\documentstyle[12pt]{article}

\begin{document}

\def\singlespace 
{\smallskipamount=3.75pt plus1pt minus1pt
\medskipamount=7.5pt plus2pt minus2pt
\bigskipamount=15pa plus4pa minus4pt \normalbaselineskip=12pt plus0pt
minus0pt \normallineskip=1pt \normallineskiplimit=0pt \jot=3.75pt
{\def\smallskip {\vskip\smallskipamount}} {\def\medskip
{\vskip\medskipamount}} {\def\bigskip {\vskip\bigskipamount}}
{\setbox\strutbox=\hbox{\vrule height10.5pt depth4.5pt width 0pt}}
\parskip 7.5pt \normalbaselines} 

\def\middlespace
{\smallskipamount=5.625pt plus1.5pt minus1.5pt \medskipamount=11.25pt
plus3pt minus3pt \bigskipamount=22.5pt plus6pt minus6pt
\normalbaselineskip=22.5pt plus0pt minus0pt \normallineskip=1pt
\normallineskiplimit=0pt \jot=5.625pt {\def\smallskip
{\vskip\smallskipamount}} {\def\medskip {\vskip\medskipamount}}
{\def\bigskip {\vskip\bigskipamount}} {\setbox\strutbox=\hbox{\vrule
height15.75pt depth6.75pt width 0pt}} \parskip 11.25pt
\normalbaselines} 

\def\doublespace 
{\smallskipamount=7.5pt plus2pt minus2pt \medskipamount=15pt plus4pt
minus4pt \bigskipamount=30pt plus8pt minus8pt \normalbaselineskip=30pt
plus0pt minus0pt
\normallineskip=2pt \normallineskiplimit=0pt \jot=7.5pt
{\def\smallskip {\vskip\smallskipamount}} {\def\medskip
{\vskip\medskipamount}} {\def\bigskip {\vskip\bigskipamount}}
{\setbox\strutbox=\hbox{\vrule height21.0pt depth9.0pt width 0pt}}
\parskip 15.0pt \normalbaselines}

\newcommand{\be}{\begin{equation}}
\newcommand{\ee}{\end{equation}}
\newcommand{\bea}{\begin{eqnarray}}
\newcommand{\eea}{\end{eqnarray}}

\begin{flushright}
IUHET-418
\end{flushright}

\begin{center}
\large {\bf Lattice simulations of the strange quark mass
and Fritzsch texture} \\ 
\vskip 2cm 
Biswajoy Brahmachari \\
\end{center}
\begin{center}
{\it Physics Department, Indiana University, Bloomington IN-47405, USA
}
\\
\end{center}
\vskip 1in
{
\begin{center}
\underbar{Abstract} \\
\end{center}
{\small 
A number of numerical simulations of lattice gauge theory have
indicated a low mass of strange quark in 100 MeV range at the scale of
$\mu=2$ GeV. In the unquenched case, which is improved over the
simulation in the quenched approximation by the inclusion of $u$ and $d$
sea quark effects, one sees a further downward trend. Here the fermion
mass 
spectrum of the Fritzsch texture is recalculated. In a single step
supersymmetric GUT with $M_X \sim 10^{16}$ GeV such values of the 
strange quark mass can be obtained for low values of $\tan \beta$. 
Experimental numbers  $m^{pole}_t = 173 \pm 6$ GeV and 
$4.1 < m_b(m_b) < 4.4$ GeV are used in this study. Since the
scenario is supersymmetric, gaugino loop diagrams contribute to the 
masses in addition to usual tree level Yukawa contributions. Upper bound
of the mixing parameter $V_{cb}$ is taken at 0.045} 
\newpage

The strange quark mass has been traditionally calculated using the
current algebra mass ratio\cite{ref1-leutwyler}
\be
{m_s \over {m_u + m_d}}=12.6 \pm 0.5 \label{eq1}
\ee
Equation (\ref{eq1}) is evaluated using  values of ($m_u + m_d$)
which are reported in  calculations of QCD finite energy sum-rules
(FESR). At the two-loop level of perturbative QCD calculations which
include non-perturbative corrections up to dimension six one
has the result\cite{ref2-dominguez}
\be
(m_u + m_d)({\rm~1~GeV})=15.5 \pm 2.0 {\rm~ MeV} \label{eq2}
\ee
For $\alpha_s(m_Z)=0.118$  results of (\ref{eq1}) and (\ref{eq2})
together lead to  
\be
m_s({\rm 1~GeV}) = 195 \pm 28 {\rm~MeV~~ or~~} m_s({\rm 2~GeV})=150 \pm 21
{\rm ~~MeV.} \label{eq3}
\ee
The ratio
\be
m_s({\rm 2~GeV})/ m_s({\rm 1~GeV})=0.769 ~~{\rm for~~}
\alpha_s(m_Z)=0.118 \label{eq4}
\ee
can be obtained by solving renormalization group 
equations\cite{ref3-barger}. A systematic uncertainty in this result
remains in reconstruction of so called `spectral function' from
experimental data of resonances. When a different functional form of the
resonance is  adopted, and three loop order perturbative QCD
theory is used one obtains\cite{ref4-bijnnes}
\be
(m_u + m_d)~({\rm 1~GeV})=12.0 \pm 2.5 ~{\rm MeV} \label{eq5}
\ee
With (\ref{eq5}) and (\ref{eq1}) one gets
\be
m_s({\rm 1 GeV})=151 \pm 32~{\rm~MeV}{\rm~~or~~}m_s({\rm 2 GeV})=116 \pm
24 ~{\rm MeV}. \label{eq6}
\ee
Again this translation from the scale of 1 GeV to the scale of 2 GeV is
obtained for the case $\alpha_s=0.118$. It has been remarked in
Ref.\cite{ref5-leutwyler-cargese} that it is indeed difficult to
account for vacuum fluctuations, or sea quark effects generated by
quarks of small masses in  perturbative QCD calculations. Thus,  
numerical simulations  of strange quark mass on a lattice becomes 
rather attractive, especially if the simulation includes virtual light
quark loop effects.

Up and down type quarks differ only in the $U(1)_{em}$ quantum numbers in
an effective theory where the gauge symmetry is $SU(3)_c \times
U(1)_{em}$. Lattice calculations in current literature have neglected
effects of $U(1)_{em}$ which distinguishes up quarks from down quarks. Let
us note that we are describing the lattice in terms of a theory at the
scale of a few GeVs where light quark masses are to be described in terms
of observables relevant to their own scales, which are 
meson masses and decay constants. Thus, lattice simulation 
determines $m_s$, $m_u + m_d \over 2$ and the lattice 
spacing $a$ using three hadronic observables. They 
can be chosen, for example,  $M_\pi, M_{K^*}$, and $f_\pi$.  
Due to the structure of equations which are to be 
fitted, the scale $a$ can also be taken as a function of some other 
observable, for example, it may be chosen as $a(M_n)$ or $a(M_\Delta)$
etc. The result depend on the choice of the observable that fits the
lattice spacing. The best choice would be the one which has minimum
experimental uncertainty and the best result would be a clever weighted
average of results from various choices. A test of the simulation
is obviously to see whether results from various observables
are statistically consistent with each other.

Next question is how do we describe quark masses when the theory
is living on a discritized lattice. Various definitions or
formalisms of quark masses on a lattice have been suggested.
Ref.\cite{ref6-gupta-E} uses the definition in terms of hopping 
parameter $\kappa$ of the lattice
\be
a~m_{bare}=log~(1+(1/{2 \kappa} - 1/{2 \kappa_c})), 
\label{eq7}
\ee 
for a Wilson-like fermion. In the continuum limit we 
have $a \rightarrow 0$, and there one gets the hopping parameter
$\kappa=\kappa_c=1/8$. A smaller hopping parameter makes the 
lattice more sticky, and fermions remain on lattice points for 
a longer time. This make them look as if they were more massive. 

There are various other formalisms of defining the mass of fermions on
the lattice such as stagerred fermions or domain wall fermions. Most 
calculations, however, use the Wilson action for various
definitions of the fermion mass. Next, to compare the result with 
experiment, one has to calculate
the $\overline{MS}$ mass at a scale $\mu$ starting from the lattice 
estimate of the bare mass (\ref{eq7}) using, for example, the mass
renormalization constant $Z_m(\mu)$ relating the lattice regularization
scheme to the continuum regularization scheme. The lattice
regularization prescription is given in Ref\cite{ref7-lepage}. The $Z_m$
constant for various formalisms such as Wilson-like or Staggered are given
in table\ref{table1} of \cite{ref6-gupta-E}. Final results of the physical
quark mass for various definitions of the fermion on a lattice differ
$O(a)$ among each other and one expects to get the same result of the
physical quark mass in the continuum limit when $a \rightarrow 0$.

Beyond the minimal lattice simulation of light quark masses
using the heavy quark effective theory, the next step would
be to incorporate sea quark effects. From the conservation of 
energy it can be understood that it is easiest to produce lightest of the
quarks virtually. Indeed such simulations have been performed. They are
termed $n_f=2$ unquenched lattice simulations. The detailed 
processes of numerical simulations are described by  
respective authors. However, we have summarized the results 
of recent studies are in table \ref{table1} in the order: 
(A)\cite{ref8-cppacs-A}, (B)\cite{ref9-gimenez-B}, 
(C)\cite{ref10-allton-C}, (D)\cite{ref11-becirevic-D}, 
(E)\cite{ref6-gupta-E}, (F)\cite{ref12-qcdsf-F}, and  
(G)\cite{ref13-gaugh-G}.

\begin{table}[hob]
\begin{center}
\[
\begin{array}{|c|c|c|c|c|}
\hline
reference & quenched & dynamical & (1/a) callibration & m_s(MeV) \\ 
\hline
A & yes & & m_\rho & 143 \pm 6 ~~\&~~ 115 \pm 2\\
B & yes & & m_{K^*} & 130 \pm 20 \\
C & yes & & m_{K^*} & 122 \pm 20 \\
D & yes & & m_{K^*} & 111 \pm 12 \\
E & yes & & m_\rho & 110 \pm 31 \\
F & yes & & m_\rho & 108 \pm 4 \\
G & yes & & 1P-1S~~splitting & 95 \pm 16 \\
\hline
A & & yes & m_\rho & 70~~\&~~80 \\
E & & yes & m_\rho &  68 \pm 19 \\
G & & yes & 1P-1S~~splitting & 54-92\\
\hline
\end{array}
\]
\end{center}
\caption{Ref. G uses 1P-1S splitting of the charmonium system
to calibrate (1/a). Reference A quotes two different results for two
sets of parameters. All results are at the scale $\mu=2$ GeV} 
\label{table1}
\end{table}

On the experimental front the bottom quark mass is in the range
\be
4.1 < m_b(m_b) < 4.4~~{\rm GeV} 
\label{eq8}
\ee
according to the review of particle physics (PDG) tables\cite{ref14-pdg}.
Theoretically, one re-expresses the bottom quark mass in terms of
parameters of the Minimal Supersymmetric Standard Model(MSSM). The tree
level contribution which is related straight to the Yukawa texture, and
the one loop contribution due to the dominant gaugino loop can be
accounted individually. Then one can write down the relation
\cite{ref15-sarid}
\bea
m_b&=& m_b^{texture} + m_b^{SUSY} \nonumber \\
&=&h_b~{V_F \over \sqrt{2}}~\cos~\beta + {m_b}~{8 \over
3}~g^2_3~{\tan \beta \over 16~\pi^2}~{m_{\tilde{g}}~\mu \over m^2_{eff}}.
\label{eq9}
\eea
Here $m_{\tilde{g}}$ is the gluino mass $\mu$ is the $\mu$ parameter
and $m_{eff}$ is averaged supersymmetry breaking mass scale. 
This paper discusses a scenario where the first term of the RHS
of (\ref{eq9}) comes from diaginalizing a Fritzsch Yukawa texture.
The second term can be estimated to be around $\pm 2$ GeV.
In the supersymmetric case it will be satisfactory if the 
Fritzsch Yukawa contribution is in the range
\be
2.1 < m_b^{texture} < 6.4 {\rm~GeV}
\label{eq10}
\ee
Next question is concerning $\tan \beta$. Supersymmetry, together with
the gauge quantum number structure of fermions demands that at least
two Higgs doublets are necessary. Hence the ratio of the VEVs of two
Higgs doublets is an unavoidable parameter given the value of the
effective four-Fermi coupling $V_F$. There are perturbative bounds
on $\tan \beta$ in the context of grand unified theories (They can be
extended to supersymmetric theories without grand unification if MSSM is
valid up to a certain high scale, say $10^{19}$ GeV). In practice the
very low-valued regions of the parameter space for $\tan \beta$ are
forbidden from perturbative considerations. See \cite{ref16-biswa-mpl} for
example. Furthermore high values of $\tan \beta$ have constraints from
charge and color breaking\cite{ref17-munoz,ref18-mar}. This is especially
true if Yukawa couplings are at the fixed point region. The intermediate
regions of $\tan \beta$ are definitely allowed. To make a safe case let
us choose the range for the purpose of this paper
\be
\tan \beta = 2-30.
\label{eq11}
\ee

Now let us focus on the texture. It has been noted that the quark mixing
angle $V_{us}$, which is a dimension-less quantity, can be thought of as
a ratio of the mass scales of flavor symmetry breaking. These symmetries
lead to the mass hierarchy between families. Phenomenologically of course,
the ratio of the masses of the first and the second generation satisfies
well the relation
\be
\tan \theta_c= \sqrt{m_d \over m_s}. 
\label{eq12}
\ee
If there are two Higgs doublets instead, (\ref{eq12}) remains
untouched as the ratio of the VEVs of the doublets cancel in the
ratio on the RHS. Thus it cannot feel $\tan \beta$. 

Suppose in a two generation case rotation angles of the up and
the down sectors are $\theta_u$ and $\theta_d$. Then
the combined quark mixing matrix $V=O_u{O_d}^\dagger$ will give
$\theta_c=\theta_u \pm \theta_d$. This observation plays
a role in the Fritzsch mass matrices. Fritzsch mass
matrices can be thought of as a set of mass matrices which 
generalizes (\ref{eq12}) to the following
form\cite{ref19-weinberg,ref20-fritzsch}
\be
\theta_c=\theta_d \pm \theta_u=~ \tan^{-1}\sqrt{m_d \over m_s} \pm
\tan^{-1}\sqrt{m_u \over m_c}.
\label{eq13}
\ee
Where $m_i$ are eigenvalues of  Fritzsch mass matrices. 
In the three generation case Fritzsch 
textures for up and down sectors are given by
\be
M_U=\pmatrix{0 & a e^{i r} & 0 \cr
             a e^{i r^\prime} & 0 & b e^{ i h} \cr
             0 & b e^{i h^\prime} & c e^{i q} }
~~
M_D=\pmatrix{0 & A e^{i R} & 0 \cr
             A e^{i R^\prime} & 0 & B e^{ i H} \cr
             0 & B e^{i H^\prime} & C e^{i Q} }.
\label{eq14}
\ee
The phases $r,R,r^\prime,R^\prime,h,H,h^\prime,H^\prime,q,Q$ can be
absorbed in the redefinition of quark fields and individual mass
matrices can be made real. However the weak charge changing current
consists of (couples to) gauge eigenstates. Furthermore the structure
of the weak charge changing current is $\psi^U_L \gamma^\mu \psi^D_L$.
Consequently the weak mixing matrix must contain some combination of
phases. It can be shown that residual phases of the weak mixing 
matrix are
\bea
\sigma & = & (r-R)-(h-H)-(h^\prime-H^\prime)+(q-Q) \nonumber\\
\tau   & = & (r-R)-(h^\prime-H^\prime).
\label{eq15}
\eea
when the weak mixing matrix is expressed as
\be
O_U \pmatrix{1 & 0 & 0 \cr
             0 & e^{ i \sigma} & 0 \cr
             0 & 0 & e^{i \tau} } {O_D}^{-1}.
\label{eq16}
\ee
In (\ref{eq16}) $O_U$ and $O_D$ diagonalizes ${\cal M}_U$ and ${\cal M}_D$
which are precisely those in (\ref{eq14}) but in the limit when all the 
phases vanish. Which are very simply
\be 
{\cal M}_U=\pmatrix{0 & a & 0 \cr
             a  & 0 & b  \cr
             0 & b  & c}
~~;~~
{\cal M}_D=\pmatrix{0 & A  & 0 \cr
             A  & 0 & B  \cr
             0 & B  & C}.
\label{eq17}
\ee
Let us consider the limit $m_u << m_c << m_t$ and $m_d << m_s << m_b$.
We can approximately re-write (\ref{eq17}) as following 
\be
{\cal M}_U=\pmatrix{0 & a & 0 \cr
                    a & -m_c & 0 \cr
                    0 & 0 & m_t}
~~;~~
{\cal M}_D=\pmatrix{0 & A  & 0 \cr
             A  & -m_d & 0  \cr
             0 & 0  & m_t}.
\label{eq18}
\ee
Now we easily see that ${\cal M}_U$ can be diagonalized by the rotation
\be
O_U=\pmatrix{\cos \theta^u_1 & \sin \theta^u_1 & 0 \cr
             -\sin \theta^u_1 & \cos \theta^u_1 & 0 \cr
              0 & 0 & 1 }
    \pmatrix{1 & 0 & 0 \cr
             0 & \cos \theta^u_2 & \sin \theta^u_2 \cr
             0 & -\sin \theta^u_2 & \cos \theta^u_2 }
\label{eq19}
\ee
where $\tan \theta^u_1= \sqrt{m_u \over m_c }$ and $\tan \theta^u_2=
\sqrt{m_c \over m_t }$. ${\cal M}_D$ can be diagonalized similarly.
In the limit of strongly hierarchical eigenvalues one may use the
approximation 
\be
\cos \theta^u_i \sim 1 ~~~~ \cos \theta^d_i \sim 1. 
\label{eq20}
\ee
Furthermore let us define the quantities
\bea
\mu_1=\sqrt{m_u \over m_c} ~~~~
\mu_2=\sqrt{m_c \over m_t} ~~~~
\nu_1=\sqrt{m_d \over m_s} ~~~~
\nu_2=\sqrt{m_s \over m_b} 
\label{eq21}
\eea
Then the mixing matrix (\ref{eq16}) takes the following form
\be
\pmatrix{ 1 
        & - \nu_1 + \mu_1 e^{ i \sigma} 
        & \mu_1 ( \nu_2 e^{ i \sigma } - \mu_2 e^{ i \tau }) \cr
           - \mu_1 + \nu_1 e^{ i \sigma} 
        & \mu_1 \nu_1 + \mu_2 \nu_2 e^{ i \sigma } + e^{ i \sigma} 
        & \nu_2 e^{ i \sigma} - \mu_2 e^{ i \tau}  \cr
           \nu_1 ( \mu_2 e^{ i \sigma } - \nu_2 e^{ i \tau }) 
        & \mu_2 e^{ i \sigma} - \nu_2 e^{ i \tau}
        & \mu_2 \nu_2 e^{i \sigma} + e^{ i \tau}
          }.
\label{eq22}
\ee
A detailed derivation of these relations are given in 
Ref\cite{ref21-shin}. A cancelation among two terms in the expression for
$V_{cb}$ in (\ref{eq22}) is needed. To achieve this one can choose
\bea
\sigma \sim \tau \sim -{\pi \over 2}
~~{\rm this~~gives}~~ 
V_{cb}  \sim  \sqrt{m_s \over m_b} - \sqrt{ m_s \over m_t}.
\label{eq23}
\eea
It is easy to check that (\ref{eq23}) makes the top quark mass too light
to be experimentally true. Thus, it is worth asking the question whether
if the Fritzsch texture were valid at the GUT scale instead, in other
words, if the flavor symmetries were exact only above the GUT scale, could
a miracle of renormalization group evolution of the masses and mixing
angles make the Fritzsch relations valid at low energy\cite{ref22-shafi}.
Here we study their idea.

The renormalization of the full $3 \times 3$ complex Yukawa matrices 
and their renormalization up to the GUT scale $M_X=10^{16}$ GeV is
what follows. Let us set our notations of  mixing angles, the
non-removable phase and eigenvalues of the Yukawa matrices. We adopt the
parameterization\cite{ref23-naculich} 
\be
\pmatrix{s_1 s_2 c_3 + c_1 c_2 e^{ i \phi}  & c_1 s_2 c_3 - s_1 c_2 e^{ i
\phi} & s_2 s_3 \cr
s_1 c_2 c_3 - c_1 s_2 e^{ i \phi} & c_1 c_2 c_3 + s_1 s_2 e^{ i \phi} & 
c_2 s_3 \cr
-s_1 s_3 & - c_1 s_3 & c_3 }.
\label{eq24}
\ee
There is a detailed proof in Ref.\cite{ref23-naculich} that in this
parameterization \underbar{eigenvalues} $y_i$ of the Yukawa textures,
three CKM mixing angles and the CP violating phase $\phi$ satisfy
the renormalization group equations 
\bea
&& 16 \pi^2 {d \over dt} \phi = 0~~, \nonumber \\
&& 16 \pi^2 {d \over dt} \ln \tan \theta_1  = -y^2_t \sin^2 \theta_3~~,
\nonumber\\
&&16 \pi^2 {d \over dt} \ln \tan \theta_2  = -y^2_b \sin^2 \theta_3~~,
\nonumber\\
&&16 \pi^2 {d \over dt} \ln \tan \theta_3  = -y^2_t -y^2_b~~, 
\nonumber\\
&&16 \pi^2 {d \over dt} \ln y_u = 
-c^u_i g^2_i + 3 y^2_t + y^2_b \cos^2 \theta_2 \sin^2 \theta_3~~,
\nonumber\\
&&16 \pi^2 {d \over dt} \ln y_c =
-c^u_i g^2_i + 3 y^2_t + y^2_b \sin^2 \theta_2 \sin^2 \theta_3~~, 
\nonumber\\
&&16 \pi^2 {d \over dt} \ln y_t =
-c^u_i g^2_i + 6 y^2_t + y^2_b cos^2 \theta_3~~,  
\nonumber \\
&&16 \pi^2 {d \over dt} \ln y_d =
-c^d_i g^2_i + y^2_t \sin^2 \theta_1 \sin^2 \theta_3 + 3 y^2_b +
y^2_\tau~~, 
\nonumber \\
&&16 \pi^2 {d \over dt} \ln y_s =
-c^d_i g^2_i + y^2_t \cos^2 \theta_1 \sin^2 \theta_3 + 3 y^2_b + y^2_\tau
~~,\nonumber \\
&&16 \pi^2 {d \over dt} \ln y_b =
-c^d_i g^2_i + y^2_t \cos^2 \theta_3 + 6 y^2_b + y^2_\tau~~,
\nonumber \\
&&16 \pi^2 {d \over dt} \ln y_e =
-c^e_i g^2_i + 3 y^2_b + y^2_\tau~~, 
\nonumber \\
&&16 \pi^2 {d \over dt} \ln y_\mu =
-c^e_i g^2_i + 3 y^2_b + y^2_\tau~~, 
\nonumber \\
&&16 \pi^2 {d \over dt} \ln y_\tau =
-c^e_i g^2_i + 3 y^2_b + 4 y^2_\tau~~.
\label{eq25}
\eea
Multipliers $c^i_j$ of gauge couplings in (\ref{eq18}) are
well known. They can be found in \cite{ref23-naculich}.
We have solved these one-loop equations numerically using Mathematica
NDSolve subroutine. The flow chart follows this line. Taking all 
experimentally possible values of the  eigenvalues but only 
central values of the angles at low energy, we have evolved the set to
the GUT scale using (\ref{eq18}). At the GUT scale  
predictions for $V_{cb}$ $V_{us}$ and $V_{ub}$ are calculated
assuming that Fritzsch relations are valid only at the GUT scale and
beyond. While translating predictions of CKM entries back to low
energy using (\ref{eq18}), we have used exact values of the
angles not central values. Thereafter we have checked whether each
individual value of masses and mixings remain within experimentally
allowed ranges. For the strange quark mass values
quoted in table\ref{table1} are used. For all other masses and
mixings the experimental values are taken from the review of particle 
physics\cite{ref14-pdg}. Our results are given in 
table\ref{table2} 

\begin{table}[htb]
\begin{center}
\[
\begin{array}{|c|c|c|c|c|c|}
\hline
\alpha_s  & \tan \beta & m_s({\rm 2~GeV})\\
\hline
0.118    &  2  &   59.90~~{\rm~MeV}     \\
0.118    &  10  &  61.52~~{\rm~MeV}     \\
0.118    &  20  &  63.05~~{\rm~MeV}     \\
0.118    &  30  &  66.90~~{\rm~MeV}     \\ 
\hline
\end{array}
\]
\end{center}
\caption{Our results are quoted for $m_t^{pole}=173$ GeV. All other
masses and mixings remain within the ranges quoted by the Review of
Particle Physics.}
\label{table2}
\end{table}

In conclusion, implications of results of $n_f=2$ unquenched lattice
simulations of the strange quark mass in the context of the Fritzsch
texture are studied. Previous calculations in this line exist in the
literature. We have two new aspects. Because the combined effect of 
charge and color breaking and perturbative unitarity of the Yukawa
couplings may rule out the large $\tan \beta$ scenario we have studied the
low $\tan \beta$ scenario. Moreover, we have included corrections of
supersymmetric origin in the study.

Supersymmetric corrections to the bottom quark mass and the low $\tan
\beta$ scenario goes hand in hand. This is in the sense that in the low
$\tan \beta$ regime Fritzsch texture demands a large Yukawa
contribution to the bottom quark mass. This is partially canceled
by the supersymmetric loop corrections. Thus, the
original Fritzsch texture is consistent with experimental data
if it holds at the GUT scale. The strange quark mass emerges
in the range 60-70 MeV at $\mu=2$ GeV for the central values of
$\alpha_s=0.118$ and $m^{pole}_t=173$ GeV. This range is consistent with
$n_f=2$ sea quark effect improved (unquenched) lattice simulations of the
strange quark mass.

We thank S. Gottlieb for discussions and M. S. Berger for lending
his Fortran routine to calculate the $\eta_i$ scaling factors
of Ref.\cite{ref3-barger}. This research was supported by 
U.S Department of Energy under the grant number DE-FG02-91ER40661.

\end{document}